\documentclass[ajl]{emulateapj}

\usepackage{natbib}
\usepackage{graphicx}
\usepackage{subfigure}
\usepackage{amsmath,amssymb}
\usepackage{color}
\usepackage{rotating}
\usepackage{hyperref}				
\hypersetup{colorlinks,%
	linkcolor=blue,%
	citecolor=blue}

\shorttitle{Quantifying the Toroidal Flux of Pre-existing Flux Ropes of CMEs}
\shortauthors{Xing et al.}
\slugcomment{Accepted for publication in ApJ}

\begin{document}
\title{Quantifying the Toroidal Flux of Pre-existing Flux Ropes of Coronal Mass Ejections}

\author{C. Xing$^{1,2}$, X. Cheng$^{1,2}$, Jiong Qiu$^{3}$, Qiang Hu$^{4}$, E. R. Priest$^{5}$, and M. D. Ding$^{1,2}$}

\affil{$^1$School of Astronomy and Space Science, Nanjing University, Nanjing, 210046, China\textup{;\\ \href{mailto:chenxing@smail.nju.edu.cn}{chenxing@smail.nju.edu.cn}; \href{mailto:xincheng@nju.edu.cn}{xincheng@nju.edu.cn}}}

\affil{$^2$Key Laboratory of Modern Astronomy and Astrophysics (Nanjing University), Ministry of Education, Nanjing 210093, China}

\affil{$^3$Physics Department, Montana State University, Bozeman, MT 59717-3840, USA}

\affil{$^4$Department of Space Science and CSPAR, The University of Alabama in Huntsville, Huntsville, AL 35805, USA}

\affil{$^5$Mathematics Institute, St Andrews University, St Andrews KY16 9SS, UK}

\begin{abstract}
In past decades, much progress has been achieved on the origin and evolution of coronal mass ejections (CMEs). In-situ observations of the counterparts of CMEs, especially magnetic clouds (MCs) near the Earth, have provided measurements of the structure and total flux of CME flux ropes. However, it has been difficult to measure these properties in the erupting CME flux rope, in particular in the pre-existing flux rope. In this work, we propose a model to estimate the toroidal flux of the pre-existing flux rope by subtracting the flux contributed by magnetic reconnection during the eruption from the flux measured in the MC. The flux by the reconnection is derived from geometric properties of two-ribbon flares based on a quasi-2D reconnection model. We then apply the model to four CME/flare events and find that the ratio of toroidal flux in the pre-existing flux rope to that of the associated MC lies in the range of 0.40--0.88. It indicates that the toroidal flux of the pre-existing flux rope has an important contribution to that of the CME flux rope and is usually at least as large as the flux arising from the eruption process for the selected events.
\end{abstract}
\clearpage

\section{INTRODUCTION}
Coronal mass ejections (CMEs) represent rapid eruptions of magnetized plasma in the solar corona, and may be observed as structures that are brighter than the background in white-light coronagraph images (\citeauthor{hundhausen1984}, \citeyear{hundhausen1984}). In a very short period of time, CMEs are accelerated from several km s$^{-1}$ to speeds that are sometimes over a thousand km s$^{-1}$ and then propagate into interplanetary space with a constant or slightly varying speed (\citeauthor{zhang2001}, \citeyear{zhang2001}). CMEs propagating in interplanetary space are also called interplanetary coronal mass ejections (ICMEs; \citeauthor{burlaga1982}, \citeyear{burlaga1982}; \citeauthor{klein1982}, \citeyear{klein1982}), some of which are termed ``magnetic clouds" (MCs) when they possess a rotation of the magnetic field (\citeauthor{burlaga1991}, \citeyear{burlaga1991}) and a decrease of proton and electron temperature (\citeauthor{richardson1995}, \citeyear{richardson1995}; \citeauthor{gosling1987}, \citeyear{gosling1987}). When MCs arrive at the Earth, they may interact with the magnetosphere and cause geomagnetic and ionospheric storms, even destroying satellite navigation and space communications (\citeauthor{gosling1993}, \citeyear{gosling1993}). 

CMEs often include a magnetic flux rope, namely, a coherent structure with all magnetic field lines twisting around a central axis, supported by internal helical bright structures within CMEs observed by the Large Angle and Spectrometric Coronagraph (LASCO) (\citeauthor{chen1997}, \citeyear{chen1997}; \citeauthor{dere1999}, \citeyear{dere1999}). It is even believed that the flux rope sometimes exists prior to the eruption, which is called the pre-existing flux rope. The evidence for a pre-existing flux rope includes (see the review by \cite{cheng2017} and references therein) filaments (e.g., \citeauthor{kuperus1974}, \citeyear{kuperus1974}; \citeauthor{priest1989}, \citeyear{priest1989}; \citeauthor{demoulin1989}, \citeyear{demoulin1989}; \citeauthor{aulanier1998}, \citeyear{aulanier1998}; \citeauthor{guo2010}, \citeyear{guo2010}), coronal cavities (e.g., \citeauthor{wang2010}, \citeyear{wang2010}), sigmoids (e.g., \citeauthor{green2009}, \citeyear{green2009}; \citeauthor{liu2010}, \citeyear{liu2010}; \citeauthor{james2018}, \citeyear{james2018}) and hot channels (e.g., \citeauthor{zhang2012}, \citeyear{zhang2012}; \citeauthor{cheng2013}, \citeyear{cheng2013}; \citeauthor{cheng2014}, \citeyear{cheng2014}).

Eruptive flares are closely related to CMEs (\citeauthor{munro1979}, \citeyear{munro1979}; \citeauthor{sheeley1983}, \citeyear{sheeley1983}; \citeauthor{webb1987}, \citeyear{webb1987}; \citeauthor{stcyr1991}, \citeyear{stcyr1991}; \citeauthor{harrison1995}, \citeyear{harrison1995}) and appear as a sudden brightening in the solar atmosphere across almost all of the electromagnetic spectrum (\citeauthor{benz2008}, \citeyear{benz2008}). The brightening often appears as two flare ribbons in the lower atmosphere, which are believed to correspond to the feet of reconnected field lines connecting opposite polarities of the magnetic field. It is worthy mentioning that the morphology of the two ribbons sometimes presents two ``J"s. The newly formed flux rope field lines are suggested to anchor at the hooked part of J-shaped ribbons, while the footpoints of flare loops mainly lie at the straight parts (\citeauthor{janvier2014}, \citeyear{janvier2014}; \citeauthor{aulanier2019}, \citeyear{aulanier2019}). Regardless of the two ribbons of eruptive flares presenting double "J"s or not, their evolution usually has two stages (\citeauthor{qiu2009}, \citeyear{qiu2009}; \citeauthor{qiu2010}, \citeyear{qiu2010}). During the first stage (zipper phase), the flare ribbons have a zipper motion, during which small patches of the chromosphere brighten on both sides of the polarity inversion line (PIL), and then they spread in a direction parallel to the PIL with a speed ranging from ten to a hundred km s$^{-1}$ and quickly form a pair of ribbons (\citeauthor{qiu2009}, \citeyear{qiu2009}; \citeauthor{qiu2010}, \citeyear{qiu2010}; \citeauthor{liu2010}, \citeyear{liu2010}; \citeauthor{cheng2012}, \citeyear{cheng2012}; \citeauthor{qiu2017}, \citeyear{qiu2017}). During the second stage (main phase), the two ribbons separate from each other in a direction perpendicular to the PIL (\citeauthor{wang2003}, \citeyear{wang2003}; \citeauthor{fletcher2004}, \citeyear{fletcher2004}; \citeauthor{qiu2004}, \citeyear{qiu2004}). The separation speed varies from tens of km s$^{-1}$ (\citeauthor{svestka1982}, \citeyear{svestka1982}; \citeauthor{wang2003}, \citeyear{wang2003}) to even 110 km s$^{-1}$ (\citeauthor{xie2009}, \citeyear{xie2009}) at the beginning to about 1 km s$^{-1}$ in the later stages (\citeauthor{wang2003}, \citeyear{wang2003}). At the same time, a row of hot arcades joining the two ribbons slowly rises along with the separation of the ribbons.

During the eruption, magnetic reconnection plays an important role in continuously building up extra flux around a pre-existing flux rope, which is named as the flux rope envelope in the following. A classical magnetic reconnection-involved model is the 2D CSHKP model, which well explains the separation motion of two-ribbon flares (\citeauthor{carmichael1964}, \citeyear{carmichael1964}; \citeauthor{sturrock1966}, \citeyear{sturrock1966}; \citeauthor{hirayama1974}, \citeyear{hirayama1974}; \citeauthor{kopp1976}, \citeyear{kopp1976}). Subsequently, \cite{shibata1999} and \cite{lin2000} interpreted CMEs by introducing a flux rope in this model. It is argued that, once the flux rope erupts, it stretches the overlying field and induces magnetic reconnection between two anti-parallel legs of the stretched field lines. The reconnection rapidly produces the twisted flux rope envelope, finally forming a CME above the reconnection site and the post-flare loops below, whose footpoints map to the two flare ribbons (\citeauthor{priest2002}, \citeyear{priest2002}). Such a model is only able to well interpret the flares with two straight ribbons. As the reconnection occurs between two legs of field lines at higher and higher altitudes, the post-flare loops rise with their footpoints separating from each other, namely the separation motion of flare ribbons. Under this 2D model, the closed fluxes that are formed during the reconnection totally go to the poloidal flux of the CME flux rope (e.g., \citeauthor{lin2004}, \citeyear{lin2004}; \citeauthor{qiu2007}, \citeyear{qiu2007}). It means that the toroidal flux of the CME flux rope is completely from the contribution of the pre-existing flux rope.

\begin{figure}
\centering
\includegraphics[height=6cm]{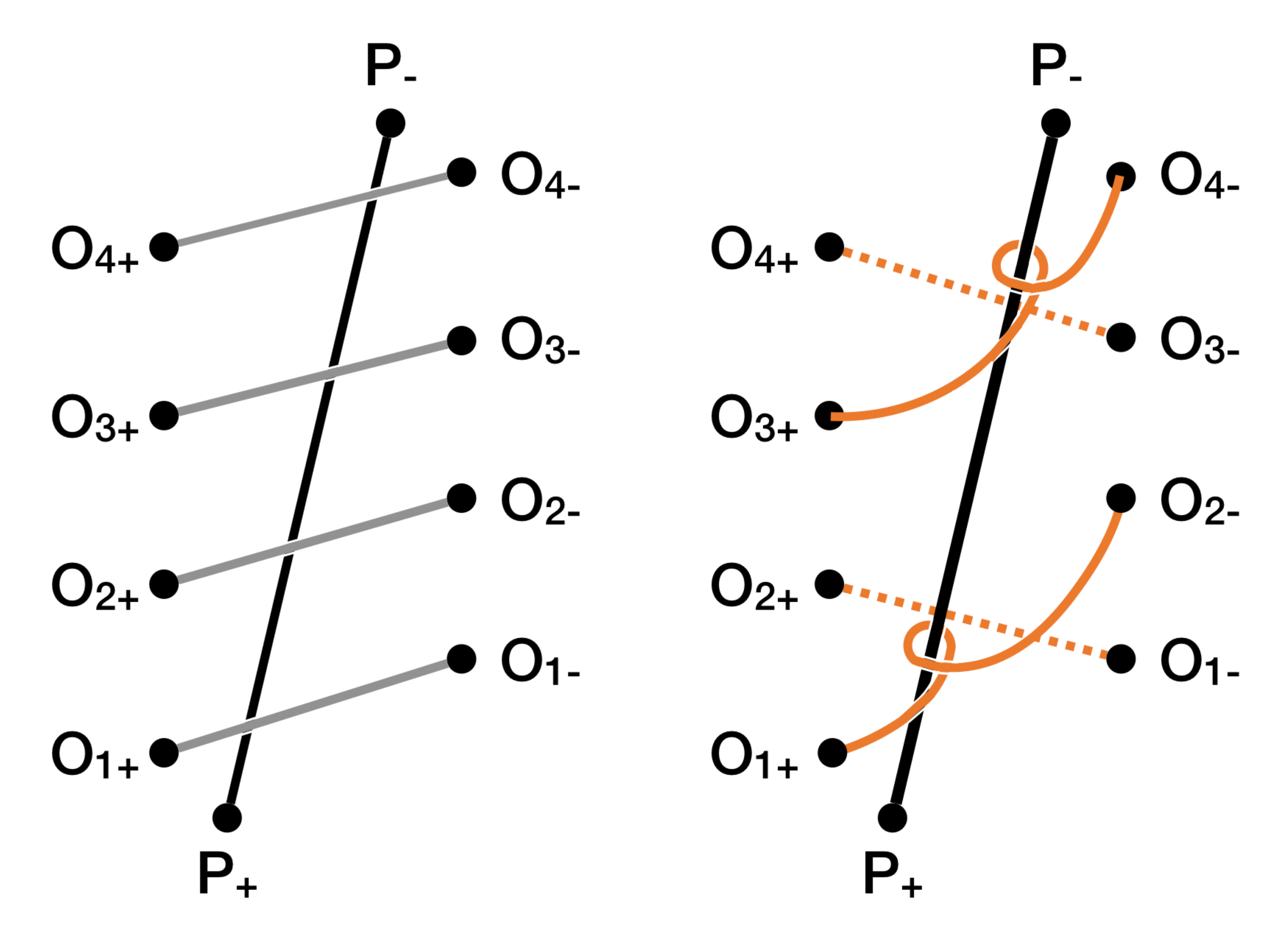}
\caption{A sketch of the quasi-2D reconnection model. Left and right panels represent magnetic structures before and after reconnection, respectively. The overlying field O$_{1+}$O$_{1-}$ (O$_{3+}$O$_{3-}$) reconnects with O$_{2+}$O$_{2-}$ (O$_{4+}$O$_{4-}$), producing a twisted field line O$_{1+}$O$_{2-}$ (O$_{3+}$O$_{4-}$) enveloping the pre-existing flux rope P$_{+}$P$_{-}$ and an arcade (presented by the dashed line) O$_{2+}$O$_{1-}$ (O$_{4+}$O$_{3-}$) lying below P$_{+}$P$_{-}$. O$_{1+}$O$_{2-}$ and O$_{3+}$O$_{4-}$ constitute the flux rope envelope. A condition for this change is that the magnetic energy of the final state is smaller than that of the initial state, so that the change is energetically possible.}
\label{fig1}
\end{figure}

However, in a real CME flux rope, the toroidal flux originates both from the pre-existing flux rope and the reconnection process because the overlying field is sheared. Thus the CSHKP model should be modified to adapt the real situation. The quasi-2D reconnection model, where the reconnection occurs between sheared overlying field lines, is an extension of the CSHKP model. It was proposed by \cite{priest2017}, also mentioned in \cite{vanballegooijen1989}, \cite{longcope2007} and \cite{green2011}. The similar process, i.e., the reconnection between sheared overlying field lines, was implemented by \cite{manchester2004}, \cite{aulanier2012} and \cite{threlfall2018} in their simulations. In the quasi-2D reconnection model, the flare ribbons are still straight and also separate from each other as the post-flare loops rise. However, different from the CSHKP model, both the flux rope envelope and the post-flare loops formed in the quasi-2D reconnection are anchored on the flare ribbons, and the newly formed twisting field lines that constitute the flux rope envelope are not self-closed any more. The envelope flux actually has two components, one is the poloidal component, the other is parallel to the axis of the magnetic flux rope, thus making a contribution to the toroidal flux of the CME flux rope. In this paper, based on the quasi-2D reconnection model, we propose a model to estimate the toroidal flux of the pre-existing flux rope for the eruptive events whose flare ribbons show primarily a separation motion, and apply it to four CME/flare events. In Section 2, we present our model. In Section 3, we quantify the model parameters and apply the method to observations. A summary and discussions are given in Section 4.

\begin{figure}
\centering
\includegraphics[height=6cm]{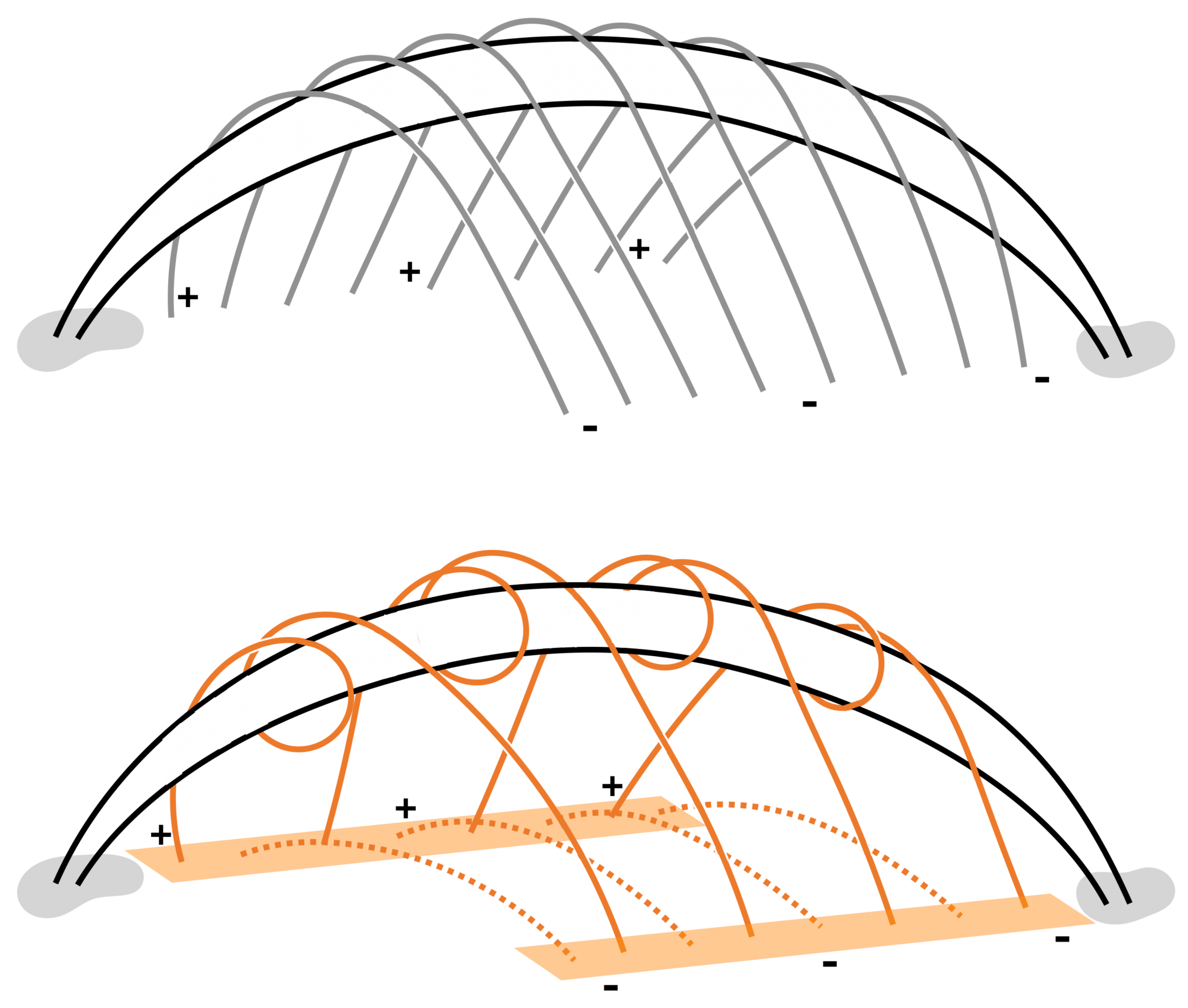}
\caption{A sketch of the quasi-2D reconnection model with a 3D view. The upper and lower panels show magnetic structures before and after the reconnection, respectively. The black curves represent the pre-existing flux rope, and the grey regions represent the two footpoints of the pre-existing flux rope. The neighbouring overlying field lines (grey solid curves in the upper panel) reconnect with each other, forming the flux rope envelope (orange solid curves in the lower panel) twisting around the pre-existing flux rope and the flare loops (orange dashed curves in the lower panel) whose footpoints map two flare ribbons (orange rhomboids in the lower panel). Compared with Figure \ref{fig1}, a difference is the number of the overlying field lines that are drawn in the figure.}
\label{fig2}
\end{figure}

\section{Model for estimating the toroidal flux of a pre-existing flux rope}
\subsection{Quasi-2D Reconnection Model}
We first introduce the quasi-2D reconnection model (Figures \ref{fig1} and \ref{fig2}) proposed by \cite{priest2017} and \cite{threlfall2018}, which grew out of earlier ideas of works by \cite{vanballegooijen1989} and \cite{green2011}. In the left panel of Figure \ref{fig1}, the line P$_{+}$P$_{-}$ represents the pre-existing flux rope and other lines represent the sheared overlying field above P$_{+}$P$_{-}$. When P$_{+}$P$_{-}$ rises up, O$_{1+}$O$_{1-}$ reconnects with O$_{2+}$O$_{2-}$, forming a field line O$_{1+}$O$_{2-}$ twisting around P$_{+}$P$_{-}$ and an arcade O$_{2+}$O$_{1-}$ lying below P$_{+}$P$_{-}$. A similar process occurs between O$_{3+}$O$_{3-}$ and O$_{4+}$O$_{4-}$, forming a twisted field line O$_{3+}$O$_{4-}$ and an arcade O$_{4+}$O$_{3-}$. O$_{1+}$O$_{2-}$ and O$_{3+}$O$_{4-}$ constitute the flux rope envelope, and reconnection does not further occur between them. Note that the lines O$_{1+}$O$_{1-}$, O$_{2+}$O$_{2-}$ and so on are parallel with each other when they indicate the connections between concentrated sources, but actually the field lines in the corona are not completely parallel since they are sheared, as seen in 3D plots of \cite{threlfall2018}. Figure \ref{fig2} shows the configuration of the flux rope before and after the quasi-2D reconnection, which is similar to that in Figure \ref{fig1} but with a 3D view. Indeed, reconnection is likely to occur when the shear is great enough and the magnetic energy of the final state is sufficiently smaller than that of the initial state that the initial state is unstable or reaches a state of nonequilibrium.

It should be mentioned that \citeauthor{priest2017}(\citeyear{priest2017}) also pointed out an extra reconnection process, which is an extension to the quasi-2D reconnection model. During the extra reconnection process, the field lines in the flux rope envelope formed by the quasi-2D reconnection could further reconnect with each other, and finally form a new envelope in which field lines twist around the pre-existing flux rope and anchor at the ends of two flare ribbons, near the footpoints of the pre-existing flux rope (such as points O$_{1+}$ and O$_{4-}$ for the flux rope envelope in the right panel of Figure \ref{fig1}). The flare ribbons are still straight, inheriting from those in the quasi-2D reconnection model. It is interesting that the new flux rope envelope and the corresponding straight flare ribbons are very similar with the simulation results by \citeauthor{aulanier2019} (\citeyear{aulanier2019}). In their work, the flare ribbons would also be mostly straight rather than J-shaped if the reconnection only occurs among the overlying field lines. This is because the corresponding flux rope envelope field lines only anchor at the tip of the hook but not the whole hook, which is close to the end of the straight part.

\begin{figure}
\centering
\includegraphics[height=6cm]{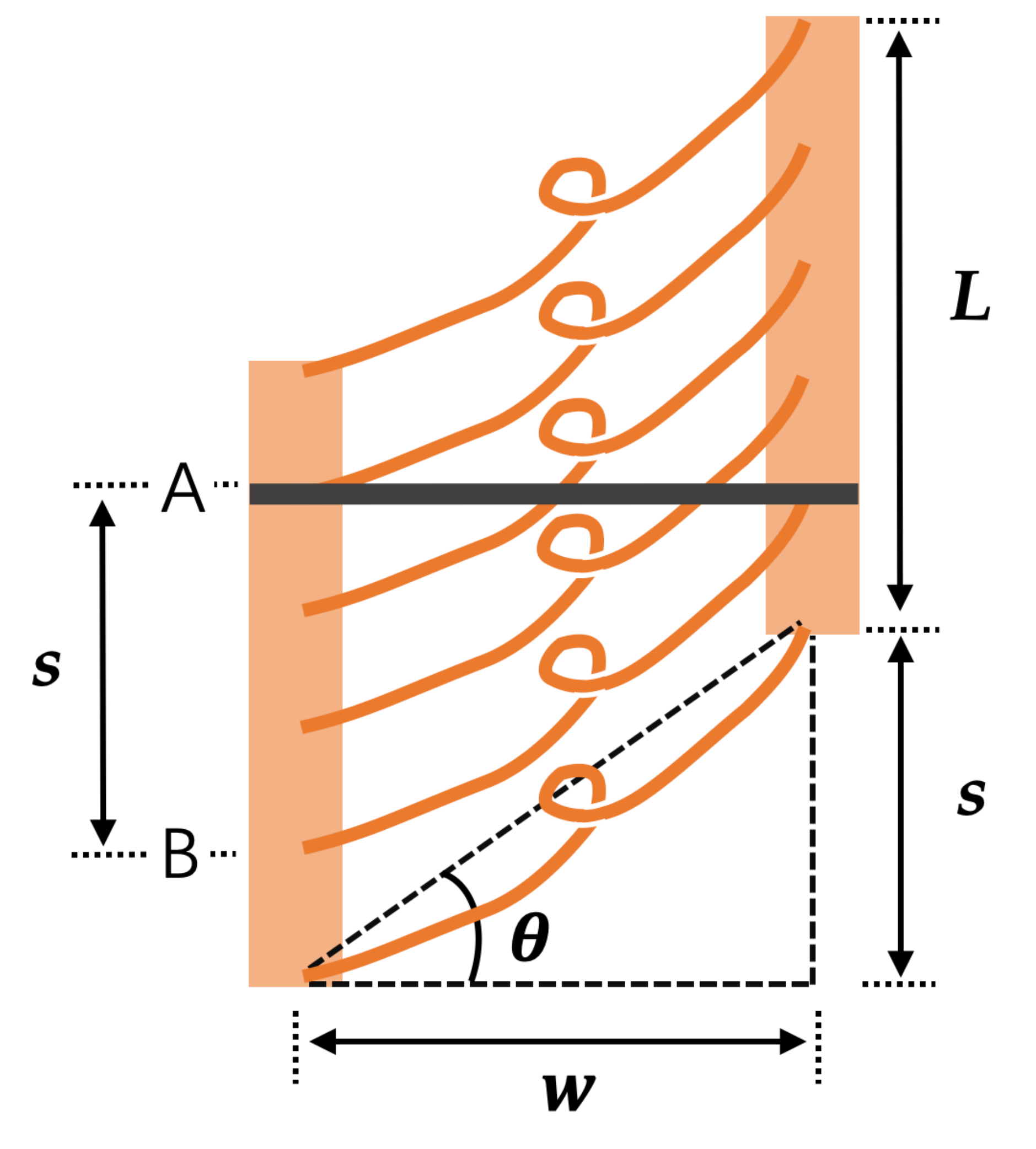}
\caption{A sketch of the flare ribbons and flux rope envelope anchored at the two flare ribbons. The orange rectangles represent the newly formed flare ribbons during a time interval $dt$, and the orange twisted lines represent the flux rope envelope formed by quasi-2D reconnection. The black solid line refers to one cross-section of the flux rope envelope. AB marks the section of the left-hand ribbon where the field lines crossing the cross-section are anchored. The length of the flare ribbons is $L$, the distance between the two flare ribbons is $w$, the inclination angle between the field line direction and the perpendicular direction is the shear angle $\theta$, and the shear distance of the two flare ribbons is $s$.}
\label{fig3}
\end{figure}

In this study, we estimate the toroidal flux of the pre-existing flux rope by subtracting the toroidal flux of the flux rope envelope from that of the whole CME flux rope. We first need to estimate the toroidal flux of the flux rope envelope which is mainly contributed by the quasi-2D reconnection but possibly reduced by an extra reconnection (see a detailed discussion in Section 2.2). However, it is almost impossible to determine how far the extra reconnection could proceed and how much it could reduce the toroidal flux of the flux rope envelope. Regardless of this, we can still approximately estimate the toroidal flux of the pre-existing flux rope based on the quasi-2D reconnection process. The reason is that the quasi-2D reconnection accounts for the primary reconnection process in the main-phase reconnection model of \cite{priest2017}. If not considering the possible reduction of the toroidal flux of the flux rope envelope by the extra reconnection, the estimated toroidal flux of the pre-existing flux rope based on the quasi-2D reconnection could be regarded as a lower limit of the actual value. In addition, although many previous studies showed that the flare ribbons may be double J-shaped rather than straight, it is demonstrated that the hooks of J-shaped ribbons are related with the field lines formed by the reconnection between the flux rope field lines and inclined ambient arcades. Nevertheless, such a reconnection does not change the toroidal flux of the flux rope (\citeauthor{aulanier2019}, \citeyear{aulanier2019}). It implies that it is reasonable to estimate the toroidal flux of the flux rope by only considering the straight part of the flare ribbons (e.g., the flare ribbons in the quasi-2D reconnection model). In the following, we adopt the quasi-2D reconnection model to quantify the toroidal flux of the pre-existing flux rope and then apply it to observations.

\begin{figure}
\centering
\includegraphics[height=6cm]{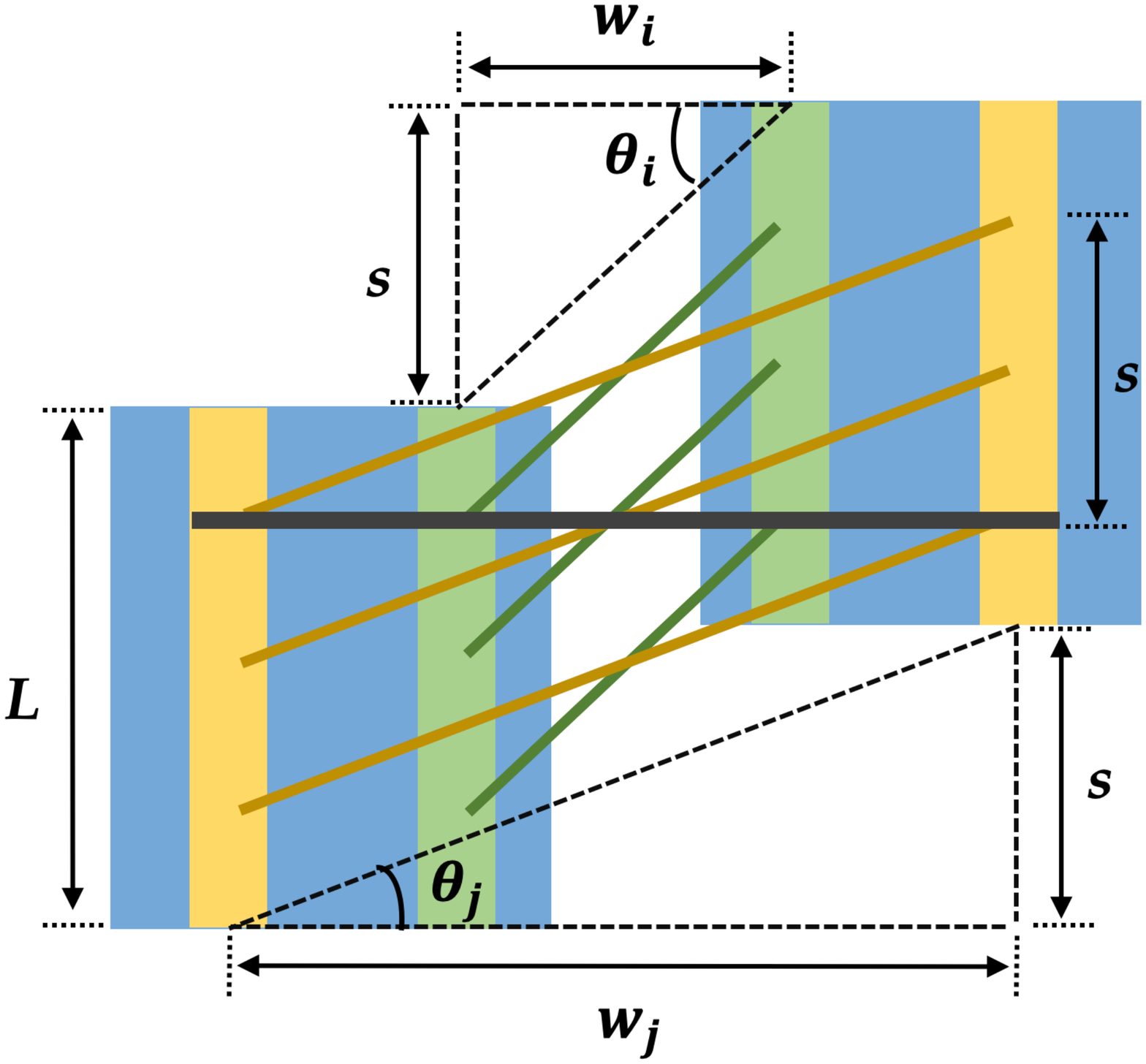}
\caption{A sketch of the evolution of the flare ribbons and the flux rope envelope. Blue rectangles represent the whole flare ribbons. Green (yellow) rectangles represent the newly formed flare ribbons during $dt_i$ ($dt_j$), and deep green (deep yellow) lines represent the flux rope envelope anchored in the ribbons (like the orange twisted lines in Figure \ref{fig3}). At the time interval $dt_i$ ($dt_j$), the distance between flare ribbons is $w_i$ ($w_j$), the flare ribbon length is $L$, the shear angle is $\theta_i$ ($\theta_j$), and the shear distance is $s$.}
\label{fig4}
\end{figure}

\begin{table*}
\renewcommand\arraystretch{1.2}
\caption{Parameters of 4 CME/flare events}
\setlength{\tabcolsep}{1.4mm}{
\begin{tabular}{ccccccccccc}
\hline
\hline
NO.&  Type$^a$&	 Source region&  Flare start time$^b$&  ICME arrival time&  $\phi_{r}$&      $\phi_{t\_MC}$&  $d/L$&  $\theta$&	    $(\phi_{t})_{0}$&  $(\phi_{t})_{0}/\phi_{t}$  \\
&     &          &              &                      &                   ($10^{21}$ Mx)&  ($10^{21}$ Mx)&  &       ($^{\circ}$)&  ($10^{21}$ Mx)&                               \\
\hline
1&    FL(X17.)&  S16E04&        2003/10/28 11:10&      2003/10/29 09:00&   23$\pm$2&       4.59$\pm$0.81&   0.20&   30$\pm$2&      1.82$\pm$0.87&     0.40$\pm$0.20              \\
2&    FL(M3.9)&	 N03E08&        2003/11/18 08:31&      2003/11/20 10:00&   3.6$\pm$0.5&     0.76$\pm$0.01&   0.24&   23$\pm$10&     0.40$\pm$0.19&     0.52$\pm$0.25              \\
3&    FL(B1.1)&	 N19W12&        2010/5/23  16:30&      2010/5/28  19:05&   0.27$\pm$0.03&   0.32$\pm$0.02&   0.19&   38$\pm$27&     0.28$\pm$0.04&     0.88$\pm$0.15              \\
4&    FL(M3.9)&	 N12W26&        2011/10/2  00:37&      2011/10/5  08:00&   1.2$\pm$0.2&     0.26$\pm$0.03&   0.33&   16$\pm$12&     0.14$\pm$0.10&     0.54$\pm$0.38              \\
\hline
\\
\end{tabular}}
\label{tab1}
\textbf{Note.}
\\$^a$ FL=flare.
\\$^b$ The flare start time refers to the time when $GOES$ X-ray flux starts to increase.
\end{table*}

\subsection{The Toroidal Flux}
In the following, based on the quasi-2D reconnection model, we first introduce the model for estimating the toroidal flux of the flux rope envelope. As shown in Figure \ref{fig3}, the orange rectangles represent a pair of newly formed flare ribbons during the time interval $dt$ and the orange twisted lines represent the flux rope envelope formed by the quasi-2D reconnection. To better show the flux rope envelope, we do not draw the flare loops, which are simultaneously formed by the quasi-2D reconnection and also anchor on the flare ribbons, and the pre-existing flux rope in the figure. It is assumed that the magnetic field is uniform at the two ribbons and the twisted lines are parallel to each other (when they indicate the connections between two footpoints rather than the real field lines). The length of flare ribbons is $L$ and the distance between the two ribbons is $w$. We refer to the direction along the flare ribbons as the parallel direction and that orthogonal to the ribbons as the perpendicular direction. The inclination angle of the flux rope envelope field lines to the perpendicular direction is the shear angle $\theta$, and the offset between the two flare ribbons along the parallel direction is the shear distance $s=w\times \textup{tan}(\theta)$. We consider any cross-section of the flux rope envelope perpendicular to the axis of magnetic flux rope as marked by the black solid line in Figure \ref{fig3}, assuming that the axis of magnetic flux rope is parallel to the PIL. The magnetic flux passing through any such cross-section is the toroidal flux of the flux rope envelope. It is obvious that only the field lines anchored between A and B can cross through that particular section. Denoting the signed magnetic flux of the flare ribbons by $\Delta\phi_{r}$ (referred to the reconnection flux for the time interval $dt$), the magnetic flux in the region between A and B is then $\Delta\phi_{r}\times s/L$ (usually $s < L$). Since during the quasi-2D reconnection, the flux should be equally allocated to the newly formed upper and lower magnetic structures (i.e., the flux rope envelope vs. the post-flare loops), only half of the flux in the region between A and B goes into the toroidal flux of the flux rope envelope. In summary, the toroidal flux contributed by flare reconnection during the time interval $dt$ is $\Delta\phi_{r}\times s/2L$.

We further take into account the temporal evolution of flare ribbons by considering a pure separation motion but no elongation of the ribbons in this quasi-2D model. In Figure \ref{fig4}, the blue rectangles represent the whole flare ribbons formed during the CME eruption. The green and yellow regions represent instantaneous flare ribbons brightened during $dt_{i}$ and $dt_{j}$, respectively. During such an evolution of the flare ribbons, the shear distance $s$ and the ribbon length $L$ are generally unchanged. Denoting the total reconnection flux during the whole flare process by $\phi_{r}$, the total toroidal flux from the flare reconnection is then $\phi_{r}\times s/2L$. Thus the total toroidal flux of the CME flux rope $\phi_t$ can be represented by:
\begin{equation}
\phi_{t}=(\phi_{t})_{0}+\frac{s}{2L}\phi_{r},
\end{equation}
where $(\phi_{t})_{0}$ represents the toroidal flux of the pre-existing flux rope.

It is obvious that, the extra reconnection occurring among the twisted field lines of the flux rope envelope formed by the quasi-2D reconnection would produce a certain flux to form new post-flare loops, which means that some toroidal flux would be removed from the flux rope envelope. Thus, the toroidal flux of the flux rope envelope should somewhat reduce after the extra reconnection. It means that the quasi-2D model would overestimate the toroidal flux of the flux rope envelope. It in turn shows that the toroidal flux of the pre-existing flux rope we derived is underestimated, thus regarded as a lower limit.

To estimate the value $(\phi_{t})_{0}$, we should measure the total toroidal flux of the CME flux rope $\phi_{t}$, the total reconnection flux $\phi_{r}$, the shear distance $s$, and the flare ribbon length $L$. In the following, we show how to determine these parameters in detail, and apply the method to four CME/flare events.

\begin{figure*}[ht]
\centering
\includegraphics[height=13cm]{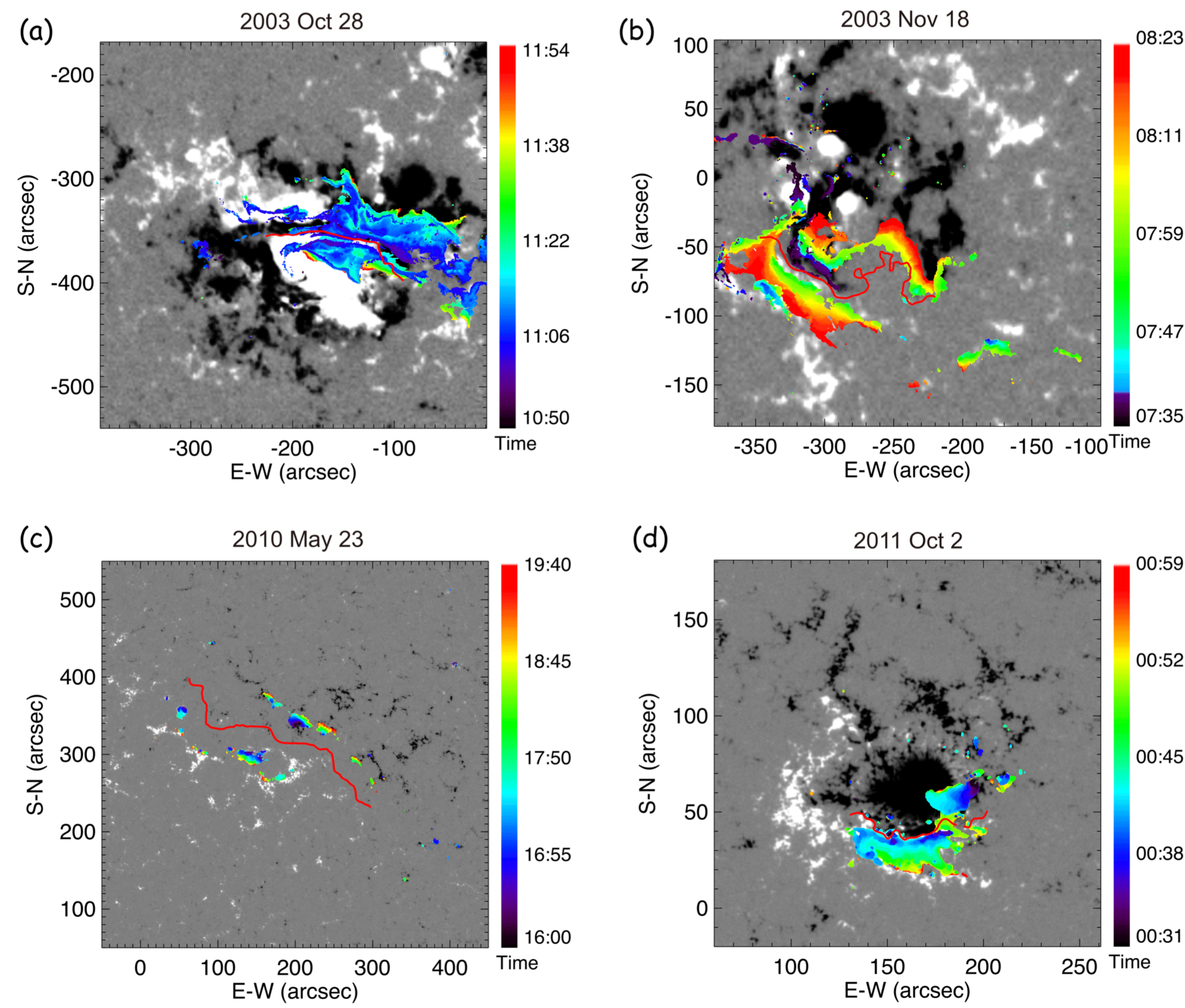}
\caption{(a) MDI line-of-sight magnetogram overlaid by two evolving flare ribbons for case 1. The evolution of the flare ribbons at the $TRACE$ 1600 $\textup{\AA}$ passband is denoted by the changing color. (b) Same as panel (a) but for case 2. (c) HMI line-of sight magnetogram overlaid by the flare ribbons at the AIA 304 $\textup{\AA}$ passband for case 3. (d) Same as panel (c) but for case 4 derived from the AIA 1600 $\textup{\AA}$ images. The intensity thresholds for the four events in panels a--d are 9, 7, 9, and 7 times the background intensity, respectively. The red curves represent the PILs for these events.}
\label{fig5}
\end{figure*}

\section{Method and its Application to Four Events}
\subsection{Event Selection}
We select events suitable for study from three lists of MC-associated CMEs provided by \cite{qiu2007}, \cite{hu2014}, and \cite{wood2017}. All events in the three lists are carefully examined. The events that are appropriate for our study should satisfy the following two criteria: (1) the source region contains a pre-existing flux rope; and (2) the flare ribbons mainly present a separation but no obvious zipper motion, and they are morphologically straight without obvious hooks. It should be noted that, observationally, there are no strictly straight ribbons since the PIL is usually curved rather than straight; the latter criterion is thus replaced by the requirement that the curved flare ribbons are mostly parallel to the PIL. We finally select four events to meet these criteria. Identifications of the association between flares, CMEs, and MCs are given by \cite{lynch2005} and \cite{qiu2007} for cases 1 \& 2, \cite{lugaz2012} and \cite{hu2014} for case 3, and \cite{wood2017} for case 4. The first three events are all accompanied with filament eruptions. The fourth event possesses a diffuse filament and a hot-channel-like structure before its eruption. These features suggest that the pre-eruptive configurations are most likely flux ropes (\citeauthor{zhang2012}, \citeyear{zhang2012}; \citeauthor{cheng2013}, \citeyear{cheng2013}; \citeauthor{cheng2014}, \citeyear{cheng2014}; \citeauthor{ouyang2017}, \citeyear{ouyang2017}). As shown in Figure \ref{fig5}, the motion pattern and morphology of flare ribbons also basically conform to the above criteria. In addition, for the four events, we do not see obvious hooked structures in the flare ribbons. It is worthy mentioning that some faint hooked structures probably exist but are undetectable. Considering that the brightness of flare ribbons is closely related to the reconnection rate, the undetectable hooks, if existing, should be produced by a weak reconnection. It means that only less magnetic flux is involved over there. Thus, neglecting the undetectable hooks will not change our results. In Table \ref{tab1}, we list the basic information of all events including the flare magnitude, location of source region, flare start time and ICME arrival time.

\subsection{Measuring Total Reconnection Flux}
The method of measuring the reconnection flux was first proposed by \cite{forbes1984} and applied to flare observations by \cite{poletto1986}, \cite{fletcher2001}, \cite{qiu2002}, \cite{asai2002}, \cite{qiu2004} and \cite{qiu2007}. Observationally, the flare ribbons are determined by identifying the regions whose intensity lies above a threshold. The reconnection flux is then calculated by integrating the magnetic flux within the flare ribbons. For our four events, the flare ribbons are determined by using the UV images observed by Transition Region and Coronal Explorer ($TRACE$; \citeauthor{handy1999}, \citeyear{handy1999}) or the EUV (or UV) images observed by the Atmospheric Imaging Assembly (AIA; \citeauthor{lemen2012}, \citeyear{lemen2012}) on board Solar Dynamics Observatory ($SDO$; \citeauthor{pesnell2012}, \citeyear{pesnell2012}). The magnetic flux in the ribbons is calculated using the photospheric magnetograms from the Michelson Doppler Imager (MDI; \citeauthor{scherrer1995}, \citeyear{scherrer1995}) on board  Solar and Heliospheric Observatory ($SOHO$) or the Helioseismic and Magnetic Imager (HMI; \citeauthor{scherrer2012}, \citeyear{scherrer2012}) on board $SDO$. In addition, the uncertainty of the total reconnection flux mainly comes from the imbalance of the flux at the positive and negative flare ribbons, as well as the uncertainty in identifying the flare ribbons.

\begin{figure*}[ht]
\centering
\includegraphics[height=18cm]{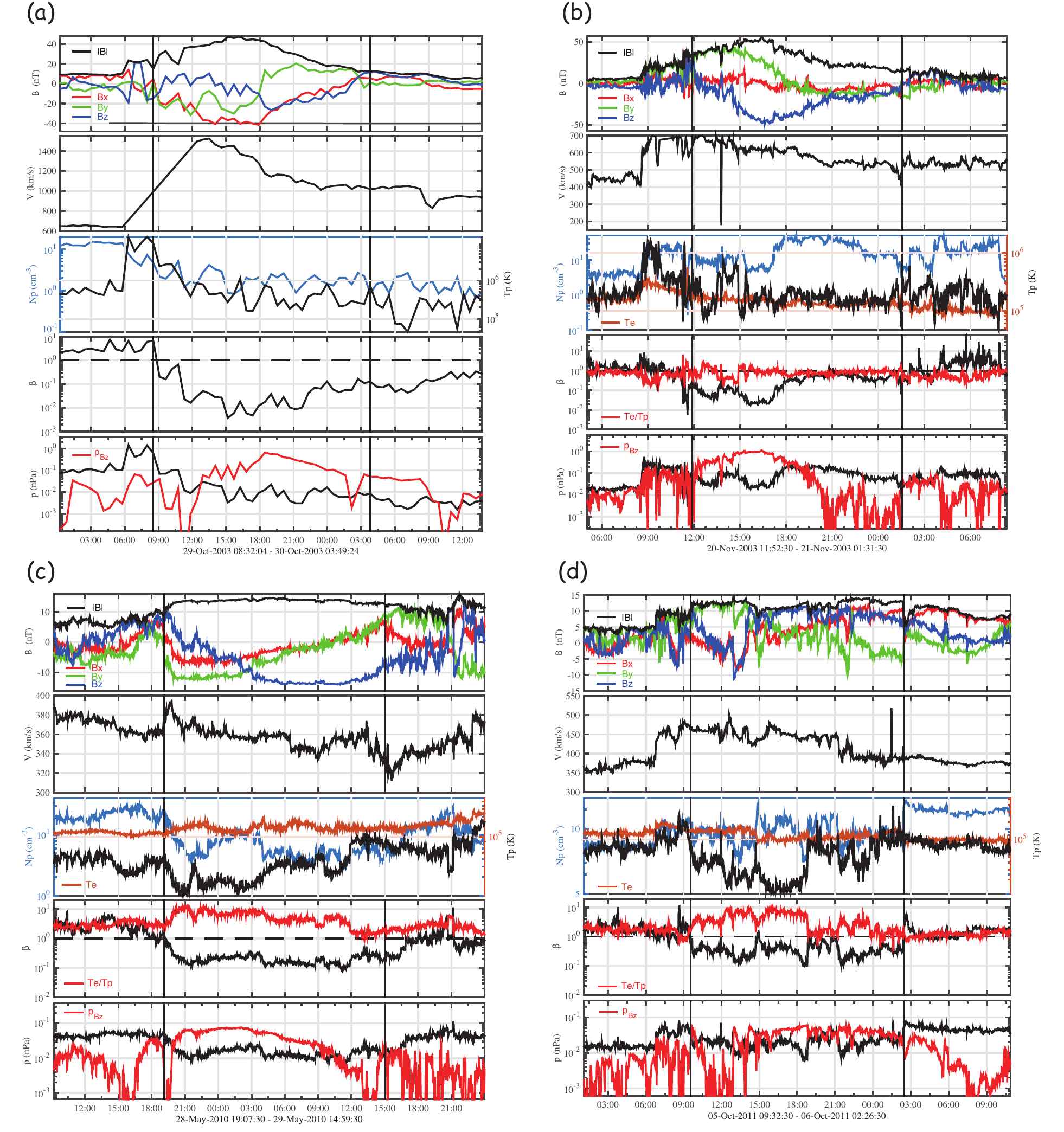}
\caption{In-situ measurement of MCs. For each panel, magnetic field strength (black) and X (red), Y (green), and Z (blue) components in GSE coordinate, plasma bulk flow speed, proton density (blue) and proton temperature (black), plasma $\beta$, and plasma pressure (black) and axial magnetic field pressure (red) are shown from top to bottom. In panels b--d, electron temperature (right axis; brown) and ratio of electron temperature to proton temperature (red) are also shown in the third and fourth sub-panels. Panels a--d are for cases 1--4.}
\label{fig6}
\end{figure*}

Figure \ref{fig5} shows the evolution of the flare ribbons for the four cases overlaid on the MDI (or HMI) line-of-sight magnetograms. For the first three cases, the total reconnection fluxes and the corresponding errors are provided by \cite{qiu2007} and \cite{hu2014}, and the ranges of intensity thresholds are also described in their papers. For the last case, the total reconnection flux and the corresponding error are calculated by ourselves, and the range of intensity thresholds is 5--8 times the intensity of the quiet Sun at the wavelength of 1600 $\textup{\AA}$. The total reconnection fluxes of the four cases and their errors are listed in Table \ref{tab1}.

\begin{figure*}[ht]
\centering
\includegraphics[height=13cm]{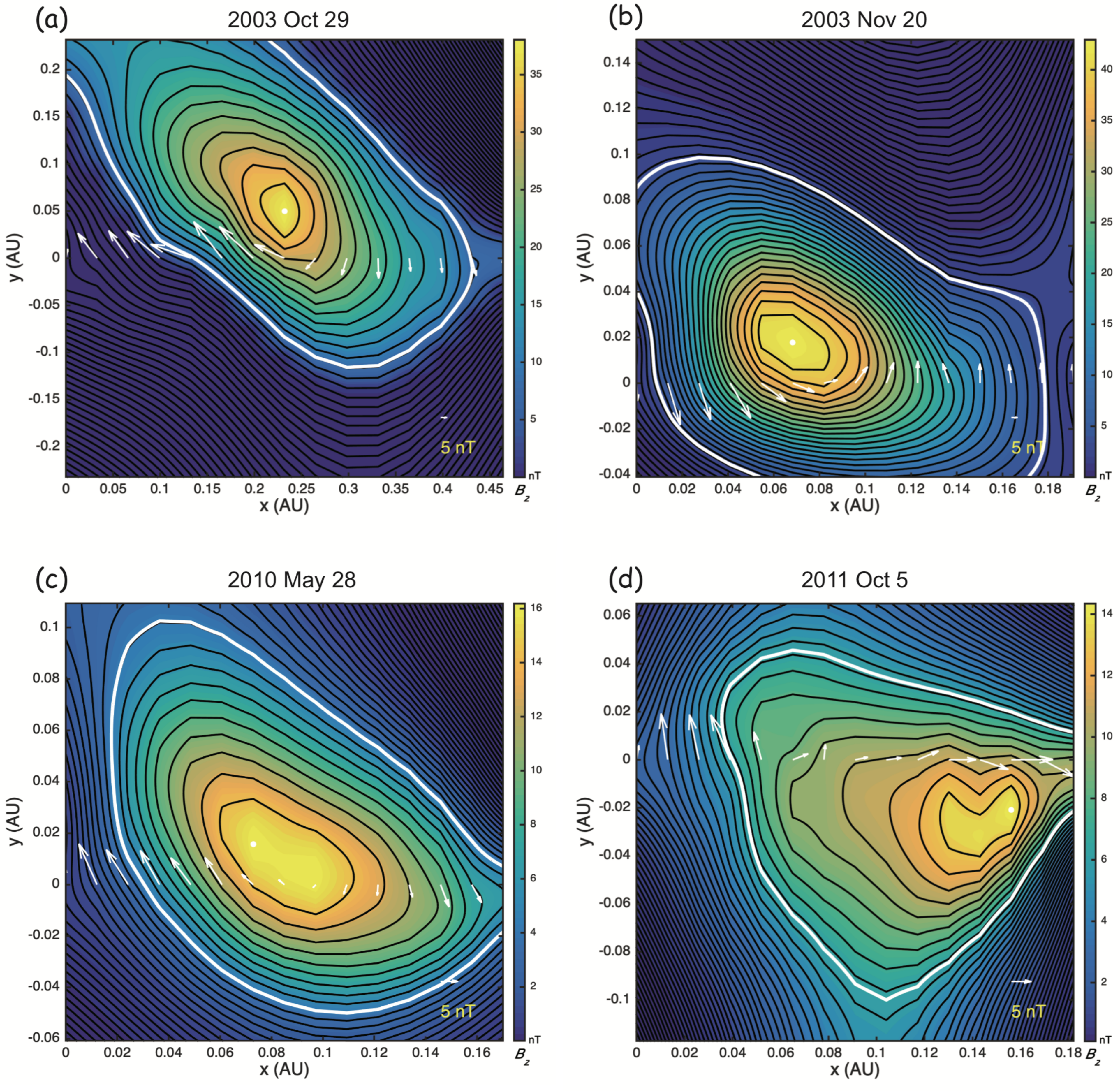}
\caption{Cross-sections of reconstructed MCs. The black curves represent the contours of $B_z$. The white point denotes the axis of MC. The boundary of MC is marked by the white curve. White arrows display the direction of magnetic field with the arrow length indicating the field strength.}
\label{fig7}
\end{figure*}

\subsection{Estimating the Toroidal Flux of CME Flux Ropes}
Assuming that MCs do not dissipate in interplanetary space, the toroidal flux of the CME flux rope $\phi_{t}$ near the Sun can be approximated by that of the MC $\phi_{t\_MC}$. We use the Grad-Shafranov (GS) reconstruction method (\citeauthor{hu2002}, \citeyear{hu2002}) to derive the 3D structure of the near-Earth MCs. It is usually believed that the GS reconstruction is able to accurately describe the MC based on a reasonable theory and initial parameters of plasma and magnetic field data. During the reconstruction process, the MC is not forcibly assumed to be force-free and its cross section is not required to be a specific shape. In the past decade, this method has been widely used to analyse the properties of MCs, especially their relation to remote-sensing observations (e.g., footpoints of the flux rope, flare ribbons, CMEs) (\citeauthor{qiu2007}, \citeyear{qiu2007}; \citeauthor{yurchyshyn2007}, \citeyear{yurchyshyn2007}; \citeauthor{hu2014}, \citeyear{hu2014}; \citeauthor{wang2017}, \citeyear{wang2017}).

The GS method assumes that the MC is a 2.5D structure and then its 2D section can be determined from the in-situ 1D data. The GS method is based on the GS function:
\begin{equation}
\frac{\partial^{2}A}{\partial x^{2}}+\frac{\partial^{2}A}{\partial y^{2}}=-\mu_{0}\frac{dP_{t}}{dA}.
\end{equation}
where the $z$-axis is along the axis of MCs such that $\partial/\partial z\approx 0$ and the $x$-axis is along the projection of the satellite trajectory in a section perpendicular to the $z$-axis. Quantity $A(x,y)\hat{z}$ is the magnetic vector potential for the transverse magnetic field $\bm{B}(x,y)$. As a single-variable function of $A$, $P_{t}(A)$ is the transverse pressure satisfying $P_{t}=p+B_{z}^{2}/2\mu_{0}$, where $p$ is the plasma pressure and $B_{z}^{2}/2\mu_{0}$ is the axial magnetic pressure. The fact that quantities $p$ and $B_{z}^{2}/2\mu_{0}$ are both functions of $A$ allows us to determine the $z$-axis. Once the $z$-axis has been determined, the distribution of $A(x,y)$ is obtained by solving the GS function using the 1D observational data and the axial magnetic field distribution $B_{z}(x,y)$ is obtained over the $A(x,y)$ solution domain. The transverse magnetic field can then be derived from $B(x,y)=(\partial A/\partial y, -\partial A/\partial x)$ and the toroidal flux is given by $\phi_{t\_MC}=\int\int B_{z}dxdy$ evaluated over the boundary $A=A_{b}$. The boundary $A=A_b$ is determined where $P_t$ transits from single-valued to multi-valued with $A$. It should also be noted that the toroidal flux obtained for MCs with this method may correspond to a lower limit. The reason is that the GS reconstruction is strictly restricted to 2.5D cases, i.e., the axis being straight, in which only the flux of the MC main body is calculated. More details on the GS reconstruction can be found in \cite{hu2002}, \cite{sonnerup2006}, and \cite{hu2017}.

For the four events in our study, we use the Advanced Composition Explorer ($ACE$) or $Wind$ data as inputs to solve the GS function. For the four cases, the temporal variations of the in-situ measurements are given in Figure \ref{fig6}, and the reconstruction results are shown in Figure \ref{fig7}. The uncertainty of the total toroidal flux $\phi_{t\_MC}$ mostly comes from the uncertainty of $z$-axis during the reconstruction process. Thus, we perform multiple reconstructions by varying the $z$-axis orientation within a certain range defined by the residue map and make an average of toroidal fluxes derived from them. The error of $\phi_{t\_MC}$ is then computed as the standard deviation. The toroidal fluxes $\phi_{t\_MC}$ and the corresponding errors for the four cases are listed in Table \ref{tab1}.

\begin{figure}
\centering
\includegraphics[height=7cm]{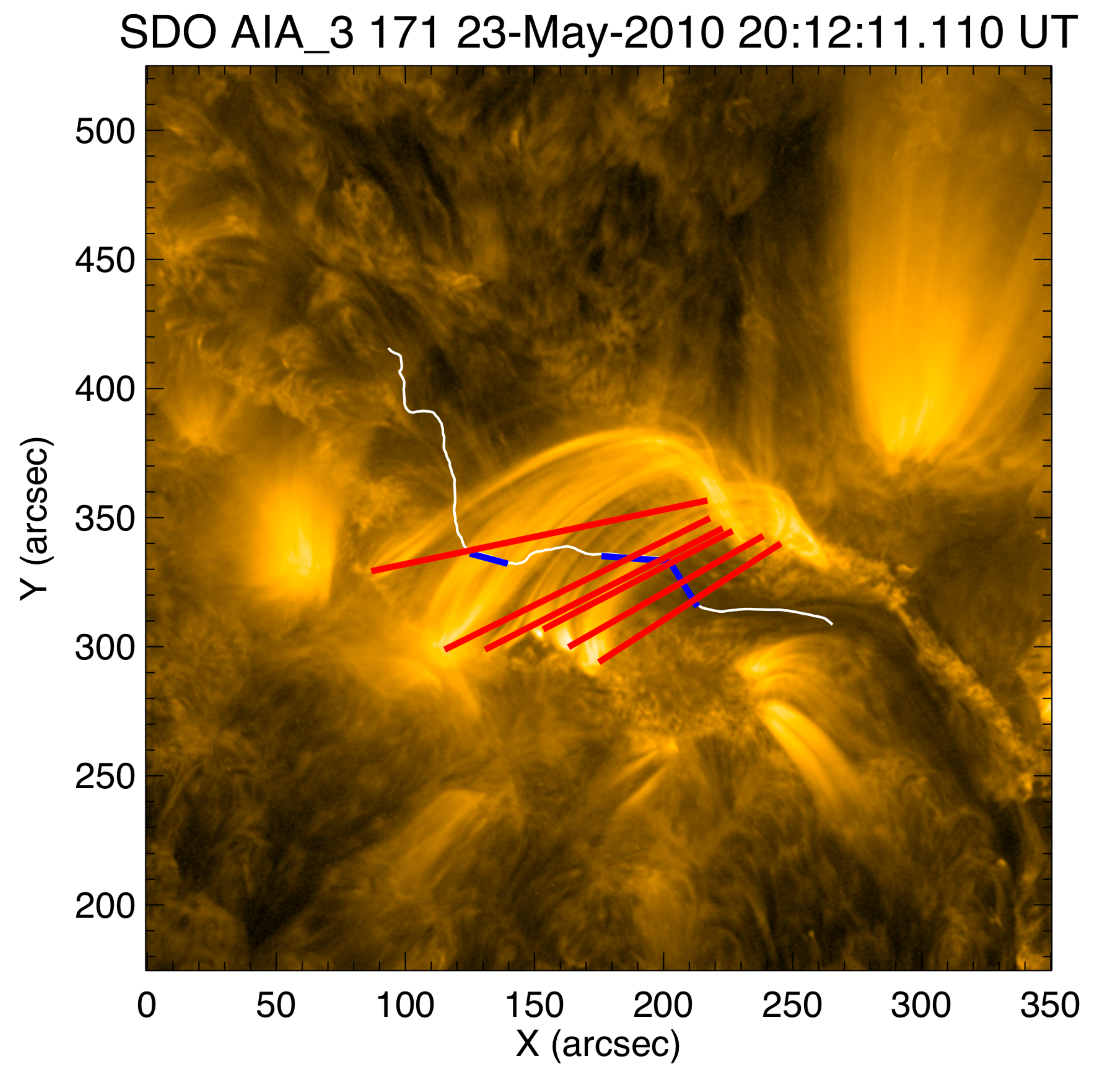}
\caption{AIA 171 $\rm \AA$ image for case 3 showing the post-flare loops. The white curves represent the PIL. The red and blue curves denote the positions of the post-flare loops and local PIL, respectively.}
\label{fig8}
\end{figure}

\subsection{Measuring Geometric Parameters of Flare Ribbons}
The parameter $s/2L$ can be derived by measuring the geometric parameters of flare ribbons. The shear distance $s$ is estimated from the ribbon distance $w$ and the shear angle $\theta$, such that $s/2L=\textup{tan}(\theta)\times w/2L$. According to our model, the parameter $s/2L$ remains constant during the flare process, which means that it can be determined through the flare ribbons at any moment. Here, we use the flare ribbons at a later phase.

The flare ribbons at the different polarities usually appear to be distinct in length. The lengths of the positive and negative ribbons are denoted by $L_1$ and $L_2$, respectively. For each polarity, we select points uniformly placed along the ribbon and measure their distances to the PIL. The average distance for each flare ribbon is denoted by $d_{1}$ and $d_{2}$, respectively. We then calculate the quantities $d/L\approx (d_{1}/L_{1}+d_{2}/L_{2})/2$ and $w/2L\approx d/L$ by assuming that the distance of each ribbon from the PIL is half the distance between the two ribbons.

The shear angle $\theta$ is estimated by measuring the inclination angle of post-flare loops to the direction perpendicular to the PIL. As two reconnected overlying field lines that we consider are very close, the orientation of the post-flare loop is almost the same as that of the envelope field line. The post-flare loops were observed by the Extreme ultraviolet Imaging Telescope (EIT; \citeauthor{delaboudini1995}, \citeyear{delaboudini1995}) on board Solar and Heliospheric Observatory ($SOHO$; \citeauthor{domingo1995}, \citeyear{domingo1995}) or the AIA on board $SDO$. Figure \ref{fig8} displays the post-flare loops for case 3 at 20:12 UT on 2010 May 23. The red lines connecting the footpoints of the post-flare loops indicate the orientations of the post-flare loops, and the blue lines are along the local PIL direction. The complementary angle of the inclination angle between red and blue lines gives the local shear angle. The final shear angle $\theta$ is calculated as the average of the local shear angles, and its error is estimated to be the standard deviation.

The geometric parameters and corresponding errors for all cases are listed in Table \ref{tab1}. It should be noted that for cases 2 and 4, one of the two ribbons is slightly irregular in shape due to the complex magnetic field distribution. In these two cases, we only measure the quantity $d/L$ for the regular ribbons.

\begin{figure}
\centering
\includegraphics[height=7cm]{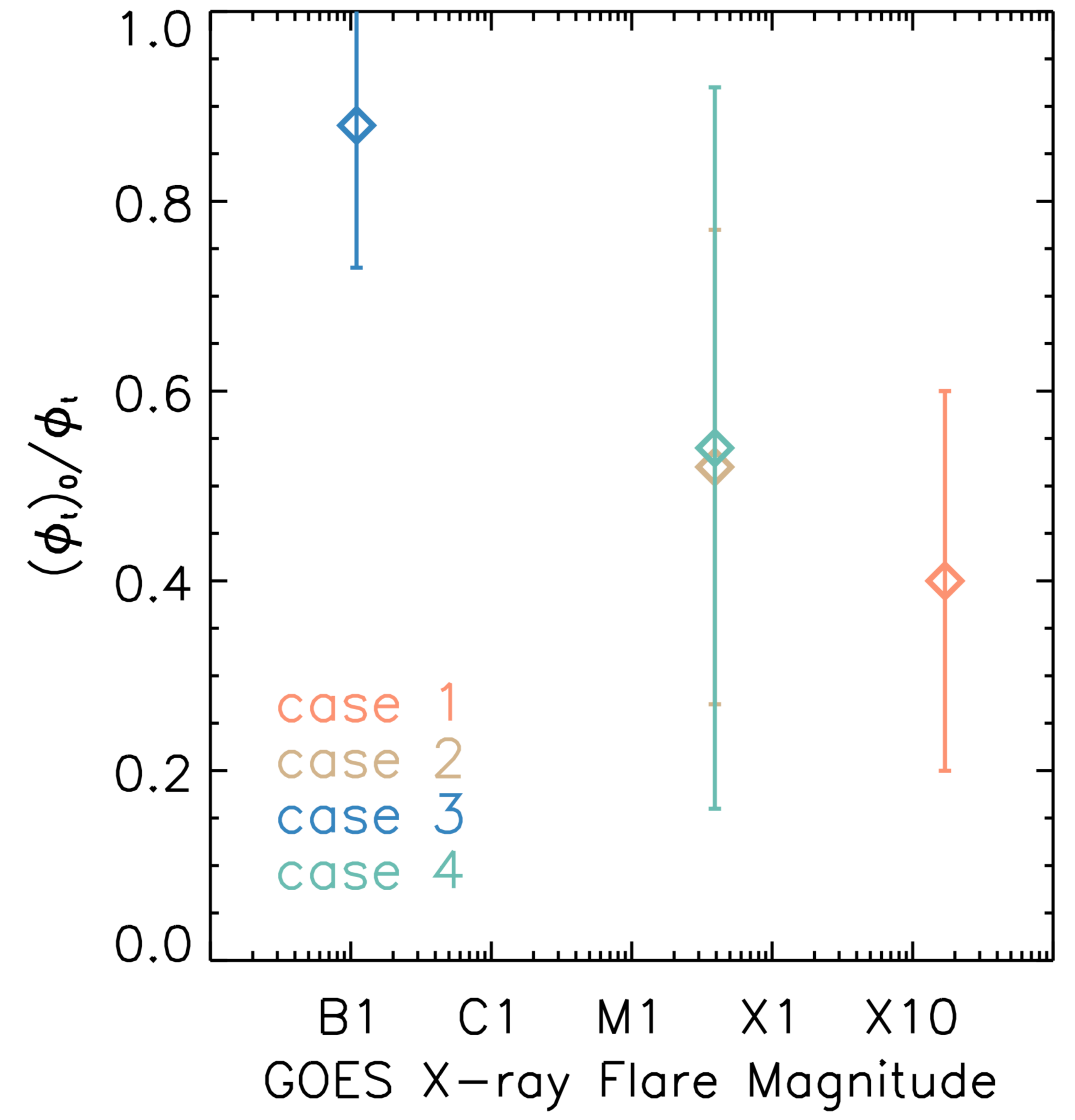}
\caption{Scatter plot of the ratio $(\phi_t)_0/\phi_t$ vs. the flare magnitude.}
\label{fig9}
\end{figure}

\subsection{Toroidal Flux of Pre-existing Flux Ropes}
Using the formula $(\phi_{t})_{0}=\phi_{t}-\phi_{r}\times s/2L$, we calculate the value $(\phi_{t})_{0}$ and derive the ratio $(\phi_{t})_{0}/\phi_{t}$, which are shown in Figure \ref{fig9} and Table \ref{tab1}. The errors of $(\phi_{t})_{0}$ and $(\phi_{t})_{0}/\phi_{t}$ primarily come from the uncertainties in quantities $\phi_{t}$, $\phi_{r}$ and $\theta$.

One finds that, the ratio $(\phi_{t})_{0}/\phi_{t}$ lies in the range of 0.40--0.88. Considering that the quasi-2D model may underestimate the toroidal flux of the pre-existing flux rope, our result shows that the pre-existing flux rope has a considerable contribution to the toroidal flux of the CMEs for these events. In addition, as shown in Figure \ref{fig9}, it seems that there is a negative correlation between the ratio $(\phi_{t})_{0}/\phi_{t}$ and the flare magnitude, which implies that the ratio may become smaller as the flare becomes stronger.

\section{Summary and Discussions}
In this paper, we quantify the toroidal flux of pre-existing flux ropes of CMEs. Based on a quasi-2D reconnection model, we propose a formula describing the variation of the toroidal flux as $\phi_{t}=(\phi_{t})_{0}+\phi_{r}\times s/2L$. We then apply it to four CME/flare events with two-ribbon structures that primarily exhibit a clear separation motion during the eruption. To derive the quantity $(\phi_{t})_{0}$ and the ratio $(\phi_{t})_{0}/\phi_{t}$, we measure the total reconnection flux and geometric parameters of the flare ribbons, and then estimate the total toroidal flux of the CME flux ropes by reconstructing the magnetic field of near-Earth MCs. For these events we study, the ratio $(\phi_{t})_{0}/\phi_{t}$ is found to be in the range of 0.40--0.88, which implies that the toroidal flux of pre-existing flux ropes provides an important contribution to that of the CMEs.

The toroidal flux of pre-existing flux ropes is closely related to the initiation of CME eruptions. The proposed initiation mechanisms for CME eruptions include reconnection-based models (\citeauthor{antiochos1999}, \citeyear{antiochos1999}; \citeauthor{moore2001}, \citeyear{moore2001}), ideal instabilities such as torus instability (\citeauthor{kliem2006}, \citeyear{kliem2006}) and kink instability (\citeauthor{torok2004}, \citeyear{torok2004}), catastrophe (\citeauthor{forbes1991}, \citeyear{forbes1991}), and force imbalance (\citeauthor{mackay2006}, \citeyear{mackay2006}). It has been suggested that under a certain overlying field, the pre-existing flux rope would experience a catastrophe when its toroidal flux increases to a critical value (\citeauthor{zhang2016}, \citeyear{zhang2016}; \citeauthor{zhang2017a}, \citeyear{zhang2017a}; \citeauthor{zhang2017b}, \citeyear{zhang2017b}). In fact, the catastrophe of the pre-existing flux rope would occur when the ratio between the toroidal flux of the pre-existing flux rope and the flux of the overlying field, $(\phi_t)_0/\phi_{ov}$ ($\phi_{ov}$ represents the flux of the overlying field), reaches a limit. In previous studies, several methods have been proposed to quantify the toroidal flux of pre-existing flux ropes, and estimate the ratio of the toroidal flux of pre-existing flux ropes to the active region flux, $(\phi_{t})_{0}/\phi_{AR}$ (here the active region flux is denoted by $\phi_{AR}$), which is closely related to the ratio $(\phi_t)_0/\phi_{ov}$. The flux rope insertion model implemented by \cite{bobra2008}, \cite{su2009} and \cite{savcheva2009} gave a value for $(\phi_{t})_{0}/\phi_{AR}$ of $10\%$--$14\%$. Applying the same strategy to several events, \cite{savcheva2012} derived the average ratio $(\phi_{t})_{0}/\phi_{AR}$ of about 36$\%$, with the smallest value of about 16$\%$. Without resorting any specific model, the cancelled flux can also be used to estimate the flux of pre-existing flux ropes. In the work by \cite{green2011}, \cite{yardley2016} and \cite{yardley2018}, the ratio of the cancelled flux to the active region flux, $\phi_c/\phi_{AR}$, ranges from $\sim$40$\%$ to $\sim$60$\%$. Since the cancelled flux is only injected partially into the pre-existing flux rope (\citeauthor{green2011}, \citeyear{green2011}), e.g., the ratio $(\phi_{t})_{0}/\phi_{c}$ could be about 60$\%$--70$\%$ (\citeauthor{savcheva2012}, \citeyear{savcheva2012}), the ratio $(\phi_{t})_{0}/\phi_{AR}$ for cases of \cite{green2011}, \cite{yardley2016} and \cite{yardley2018} can be corrected to be 20$\%$--40$\%$.

For comparison, the ratio $(\phi_{t})_{0}/\phi_{AR}$ for the cases 1, 2, and 4 in this study from active regions is 3$\%$--5$\%$, which is smaller than the ratio derived by the above two methods. Such a small ratio may be caused by the following reasons. On one hand, the toroidal flux of the MC may be underestimated by the GS reconstruction method, as discussed in Section 3.3. The flux of MCs may also decrease during their propagation in the interplanetary space (\citeauthor{dasso2007}, \citeyear{dasso2007}; \citeauthor{wang2018}, \citeyear{wang2018}). Both factors imply that we may underestimate the toroidal flux of the CME flux rope near the Sun. On the other hand, we may overestimate the toroidal flux of the flux rope envelope by using the quasi-2D reconnection model, as discussed in Section 2.2. Therefore, the toroidal flux of the pre-existing flux rope, i.e., the toroidal flux of the CME subtracted by that of the flux rope envelope, should be somewhat higher than what we have obtained here. In addition, for these three cases, the overlying field that participates in the reconnection process may be only a small part of the whole field of the active region, thus resulting in a small ratio $(\phi_{t})_{0}/\phi_{AR}$.

It should also be mentioned that we assume that the local shear angles of different post-flare loops are equally weighted when estimating the average shear angle. However, different loops are rooted in regions with different magnetic fluxes. This implies that including the weight of the flux could improve our results in principle. However, this is impractical at present since the footpoints of the post-flare loops are not well observed. We expect that observations with a higher spatial resolution can help address this issue in the future.

\begin{acknowledgements}
We are grateful to the referee for his/her constructive comments that helped improve the manuscript. We thank the $ACE$ Science Center and NASA CDAWeb for providing $ACE$ and $Wind$ spacecraft data. $SDO$ is a mission of NASA’s Living With a Star Program. $SOHO$ is a project of international cooperation between ESA and NASA. C.X., X.C. and M.D.D. are funded by NSFC grants 11722325, 11733003, 11790303, 11790300, Jiangsu NSF grants BK20170011, and ``Dengfeng B" program of Nanjing University.
\end{acknowledgements}

\end{document}